\documentclass[a4paper, 10pt, twocolumn]{article}

\usepackage[utf8]{inputenc}         
\usepackage[english]{babel}          

\usepackage{graphicx}               
\usepackage{tabularx}               
\usepackage{multirow}               
\usepackage{url}                

\usepackage[small,bf]{caption2}     
\usepackage{parskip}
\usepackage{titlesec}
\usepackage{amsmath}    
\usepackage{hyperref}
\usepackage{booktabs}
\usepackage{pdflscape}
\usepackage{listings}

\usepackage{bm}

\usepackage[svgnames]{xcolor}

\lstdefinelanguage{XML}
{
  basicstyle=\ttfamily\footnotesize,
  morestring=[b]",
  moredelim=[s][\bfseries\color{Maroon}]{<}{\ },
  moredelim=[s][\bfseries\color{Maroon}]{</}{>},
  moredelim=[l][\bfseries\color{Maroon}]{/>},
  moredelim=[l][\bfseries\color{Maroon}]{>},
  morecomment=[s]{<?}{?>},
  morecomment=[s]{<!--}{-->},
  commentstyle=\color{DarkOliveGreen},
  stringstyle=\color{blue},
  identifierstyle=\color{red}
}

\titleformat{\section}{\normalfont\large\bfseries}{\thesection}{}{}
\titleformat{\subsection}{\normalfont\large\bfseries}{\thesection}{}{}
\titleformat{\paragraph}{\normalfont\bfseries}{\theparagraph}{}{}
\titlespacing{\section}{0pt}{6pt}{-1pt}
\titlespacing{\subsection}{0pt}{3pt}{-1pt}
\titlespacing{\paragraph}{0pt}{3pt}{-1pt}

\newcolumntype{Y}{>{\centering\arraybackslash}X}    

\usepackage{tabu}

\begin{document}

\date{}                                         

\title{\vspace{-8mm}\textbf{\large
Offline coupling of segregated multi-physical simulations with consistent boundary conditions and source terms based on scattered data }}

\author{
Stefan Schoder  \\
\emph{\small TU Graz, Aeroakustik und Vibroakustik, IGTE, Inffeldgasse 18, A-8010 Graz, AT, Email: stefan.schoder@tugraz.at
} 
} \maketitle
\thispagestyle{empty}           

\section*{Abstract}

This article presents the openCFS submodule scattered data reader for coupling multi-physical simulations performed in different simulation programs. For instance, by considering a forward-coupling of a surface vibration simulation (mechanical system) to an acoustic propagation simulation using time-dependent acoustic absorbing material as a noise mitigation measure. The nearest-neighbor search of the target and source points from the interpolation is performed using the FLANN or the CGAL library. In doing so, the coupled field (e.g., surface velocity) is interpolated from a source representation consisting of field values physically stored and organized in a file directory to a target representation being the quadrature points in the case of the finite element method. A test case of the functionality is presented in the "testsuite" module of the openCFS software called "Abc2dcsvt". This scattered data reader module was successfully applied in numerous studies on flow-induced sound generation. Within this short article, the functionality, and usability of this module are described.

\section*{Implementation of the scattered data reader}
Within this contribution, we concentrate on the openCFS \cite{CFS} module \textit{scattered data reader}. The scattered data reader has several interpolation possibilities and datatype dependencies, ranging from spatially varying fields (so-called "csv") to spatially and temporally varying fields (so-called "csvt" type of scattered data reader). This csvt is a general implementation of field interpolation inside the finite element method core of openCFS and will be discussed in the following.

\subsection*{File structure and data organization}
Each point of the scattered data points is described by its coordinates (x,z,y), they are stored in the file name given in the user definition. 
A common (offline) approach to organize the input data is to use separate files. The organization of the files is done in a master file, pointing towards several subfiles. This master file contains information about the file path of the point location file and the time step value (physical time), and the path to the corresponding data file. It lists the timestamps that correspond to the data in the subsequent files. The file with the point locations contains the coordinates of the data points or locations (x,y,z). Each row typically represents a single point, and the columns correspond to the spatial dimensions in the Cartesian coordinate system (x,y,z). For each time step mentioned in the master file, there will be a separate data file. These files contain the values measured at each corresponding point location. Each data file is organized as a table, where the rows represent the points, and the columns represent different data attributes or variables. The number of points in the geometry and data file must coincide.

\subsection*{Nearest-neighbor search algorithm}
As performed in \cite{CFSDAT}, the CGAL \cite{cgal} or the FLANN \cite{flann} library may be used to conduct the k-nearest-neighbor search regarding a certain distance measure (norm). As default setting, an isotropic hypersphere-search (distances based on the Euclidean norm) is conducted. 

\subsection*{Interpolation}
An automatic conduction of the interpolation or approximation techniques to estimate the values at non-provided points is the purpose of this scattered-data reader. The interpolation of the data is based on Shepard's method 
$$
f(\bm y_k) = \frac{\bm f_{i} w_{ik}}{\delta_{ij} w_{ik}}
$$
using the linear combination\footnote{Einsteins convention of summation is used.} of the data values $\bm f_{i}$ and weights $w_{ik}$ associated to a source point $\bm x_i$
$$
w_{ik} = \frac{1}{d(\bm x_i,\bm x_k)^p} \, ,
$$
and a target point $\bm x_k$, the weighting exponent $p$ and with the distance measure $d()$, e.g., the Euclidean distances.
The interpolation methods can be easily extended to more sophisticated methods such as kriging, radial basis functions \cite{schoder2020radial,schoder2020aeroacoustic}, FEM interpolation based on triangulation \cite{CFSDAT}. 

\subsection*{Connection to the finite element method}
The finite element method (FEM) simulation discretizes the domain into finite elements, each with a quadrature point set (e.g., Gauss, Gauss-Lobatto, ...). These quadrature points serve as integration points for the numerical integration of the governing equations within each finite element. However, the original data might not align perfectly with the quadrature points, and the scattered data reader is used to provide the data here. Nearest-neighbor interpolation is a straightforward and computationally efficient method used to estimate the values at the quadrature points based on the data from the scattered points. The process can be summarized in the following steps: Find the nearest neighboring scattered data points for each quadrature point. This can be achieved by calculating the distances between the quadrature point and all the scattered points and selecting the k closest ones (FLANN or CGAL). Assign weights to the neighboring scattered data points based on their distance to the quadrature point and the interpolation method used. Generally speaking, the closer a scattered point is to the quadrature point, the higher its weight in the interpolation. Perform a weighted interpolation using a linear combination of the assigned weights and the data values of the neighboring scattered-data points. 
Repeat this process for all finite elements and quadrature points in the finite element mesh. After the scattered data reader completes the nearest-neighbor interpolation, the FEM simulation can proceed with numerical integration, solving the governing equations. The scattered data reader with nearest neighbor interpolation provides a handy tool for processing irregularly distributed data to boundary conditions or source terms of multi-physical simulations using openCFS.

\subsection*{Usage}
Within the simulation program, it can be used to interpolate field densities as source terms. In FEM simulations, the domain is discretized into finite elements, each containing a set of quadrature points. Nearest neighbor interpolation transfers data from scattered points within the domain to the quadrature points within the finite elements.

Similar to using the data for the source terms, it can be used to prescribe inhomogeneous boundaries. In FEM simulations, boundaries are discretized using quadrature points to compute boundary integrals. Nearest neighbor interpolation is employed to obtain values at the quadrature points on the boundary from the scattered data points.

In both variants, the nearest neighbor interpolation is applied to transfer data from scattered points to quadrature points within the domain or on the boundary. This allows the FEM simulation to work with the interpolated values and perform computations effectively. 

\subsection*{Setup}
As illustrated in the "Abc2dcsvt" testsuite example of openCFS \cite{CFS}, first register a scattered data reader by the id and the file-name
\begin{lstlisting}[language=XML]
...
    <scatteredData>
      <csvt fileName="./csvt/data.descrip" id="myCSVT">
        <stepValues col="0"/>
        <stepFiles col="1"/>
        <coordinates>
          <comp dof="x" col="0"/>
          <comp dof="y" col="1"/>
<!--          <comp dof="z" col="2"/>-->
        </coordinates>
        <quantity name="scatter" id="acouPot" knnLib="Flann">
          <comp col="0"/>
          <!--          <comp dof="y" col="1"/>-->
          <!--          <comp dof="z" col="2"/>-->
        </quantity>
      </csvt>
    </scatteredData>
...
\end{lstlisting}
The "stepValues" parameter "col" defines the line of the file path to the file containing the source point locations. With "stepFiles" and "col" one defines the line of the master-file where each physical time is stored (comma-seperated) followed by the file path to the data files in the same line. For each time step a new line is added to the file. The structure of the source point location file is defined within the xml-tag "coordinates". With the xml-tag "quantity", a scattered field is defined by the "id" and the nearest-neighbor search library "Flann"/"Cgal". The "id" is used later for using the "scatteredData" for boundaries or source terms using the same "id" as "quantityId" (see the second xml snippet). 

In multiphysics simulations, different physical processes often interact with each other at the boundaries. These interactions may involve gradients, fluxes, or other boundary conditions that vary spatially and temporally \cite{schoder2019hybrid}. The scattered data reader can efficiently organize and interpolate the inhomogeneous boundary terms to the corresponding boundary elements, allowing for accurate representation and simulation of complex boundary conditions. Then, one uses the scattered data for arbitrary boundary conditions within an openCFS simulation file.
\begin{lstlisting}[language=XML]
...
          <normalVelocity name="dombnd">
            <scatteredData quantityId="acouPot"/>
          </normalVelocity>
...
\end{lstlisting}
The same can be done for any field. In this case, the openCFS community recommends using openCFS-Data \cite{CFSDAT}.

\section*{Application and Summary}
This article presents a scattered data reader for connecting segregated multi-physical simulations. For instance, 
a hybrid aeroacoustic workflow can be realized efficiently \cite{schoder2022aeroacoustic,schoder2022cpcwe,schoder2021application,tieghi2023machine,schoder2022dataset,lenarcic2015numerical}. The methods' strength is leveraged, as the prescription of inhomogeneous boundary conditions for calculating the Helmholtz decomposition  \cite{schoder2020postprocessing,schoder2020postprocessing2,schoder2019helmholtz,schoder2022post} is possible. As performed recently, automotive OEMs use many software packages; the method supports an easy-to-integrate and easy-to-maintain interface \cite{freidhager2021simulationen,weitz2019numerical,maurerlehner2022aeroacoustic}. 

In Multiphysics simulations, it is often computationally efficient to solve individual physical processes separately (segregated) and then exchange information iteratively until convergence is achieved. The scattered data reader's capability to handle irregularly distributed data allows each physical process to be simulated independently with its specific boundary and source data, ensuring accurate coupling between the domains. Consider a multiphysics simulation involving fluid-structure interaction (FSI), where the fluid and structural domains are simulated independently and exchange information at the interface in each iteration. The scattered data reader can facilitate this information exchange by handling the irregularly distributed data at the interface, ensuring consistent and accurate coupling between the fluid and structural simulations.


In this article, the scattered data reader module of openCFS was presented. The method can use CGAL or FLANN libraries for the geometrical operations. In doing so, the reader provides an easy-to-integrate and easy-to-maintain interface for segregated multi-physical simulations using openCFS. In conclusion, the scattered data reader's ability to organize and interpolate inhomogeneous boundary terms and source values is paramount in multiphysics simulations. Its capability to handle irregularly distributed data points allows for an accurate representation of complex boundary conditions and sources, enabling researchers and engineers to perform segregated simulations of coupled physical phenomena efficiently and accurately. Finally, an automated method for predicting outliers in the offline connection may be valuable to consider \cite{schoder2022error}. The combination with the automated outlier detection makes the scattered data reader a valuable and robust tool in various scientific and engineering domains where multiphysics simulations are essential for gaining insights and making informed decisions. 

\bibliographystyle{abbrv}
\bibliography{references} 

\begin{thebibliography}{10}

\bibitem{freidhager2021simulationen}
C.~Freidhager, P.~Maurerlehner, K.~Roppert, A.~Wurzinger, A.~Hauser,
  M.~Heinisch, S.~Schoder, and M.~Kaltenbacher.
\newblock Simulationen von str{\"o}mungsakustik in rotierenden bauteilen zur
  entwicklung von antriebskonzepten der autos der zukunft.
\newblock {\em e \& i Elektrotechnik und Informationstechnik}, 138(3):212--218,
  2021.

\bibitem{lenarcic2015numerical}
M.~Lenarcic, M.~Eichhorn, S.~Schoder, and C.~Bauer.
\newblock Numerical investigation of a high head francis turbine under steady
  operating conditions using foam-extend.
\newblock In {\em Journal of Physics: Conference Series}, volume 579, page
  012008. IOP Publishing, 2015.

\bibitem{maurerlehner2022aeroacoustic}
P.~Maurerlehner, S.~Schoder, J.~Tieber, C.~Freidhager, H.~Steiner, G.~Brenn,
  K.-H. Sch{\"a}fer, A.~Ennemoser, and M.~Kaltenbacher.
\newblock Aeroacoustic formulations for confined flows based on incompressible
  flow data.
\newblock {\em Acta Acustica}, 6:45, 2022.

\bibitem{flann}
M.~Muja and D.~G. Lowe.
\newblock {\em {FLANN}: Fast library for approximate nearest neighbor}.
\newblock {M.~Muja}, {1.8.4} edition, 2013.

\bibitem{schoder2022cpcwe}
S.~Schoder.
\newblock cpcwe--perturbed convective wave equation based on compressible
  flows.
\newblock {\em arXiv preprint arXiv:2209.11410}, 2022.

\bibitem{schoder2022dataset}
S.~Schoder and F.~Czwielong.
\newblock Dataset fan-01: Revisiting the eaa benchmark for a low-pressure axial
  fan.
\newblock {\em arXiv preprint arXiv:2211.12014}, 2022.

\bibitem{schoder2020radial}
S.~Schoder, C.~Junger, K.~Roppert, and M.~Kaltenbacher.
\newblock Radial basis function interpolation for computational aeroacoustics.
\newblock In {\em AIAA AVIATION 2020 FORUM}, page 2511, 2020.

\bibitem{schoder2019hybrid}
S.~Schoder and M.~Kaltenbacher.
\newblock Hybrid aeroacoustic computations: State of art and new achievements.
\newblock {\em Journal of Theoretical and Computational Acoustics},
  27(04):1950020, 2019.

\bibitem{schoder2019helmholtz}
S.~Schoder, M.~Kaltenbacher, and K.~Roppert.
\newblock Helmholtz's decomposition applied to aeroacoustics.
\newblock In {\em 25th AIAA/CEAS Aeroacoustics Conference}, 2019-2561.

\bibitem{schoder2022aeroacoustic}
S.~Schoder, M.~Kaltenbacher, {\'E}.~Spieser, H.~Vincent, C.~Bogey, and
  C.~Bailly.
\newblock Aeroacoustic wave equation based on pierce's operator applied to the
  sound generated by a mixing layer.
\newblock In {\em 28th AIAA/CEAS Aeroacoustics 2022 Conference}, page 2896,
  2022.

\bibitem{schoder2022error}
S.~Schoder, F.~Kraxberger, S.~Falk, A.~Wurzinger, K.~Roppert, S.~Kniesburges,
  M.~D{\"o}llinger, and M.~Kaltenbacher.
\newblock Error detection and filtering of incompressible flow simulations for
  aeroacoustic predictions of human voice.
\newblock {\em The Journal of the Acoustical Society of America},
  152(3):1425--1436, 2022.

\bibitem{schoder2022post}
S.~Schoder, E.~Museljic, F.~Kraxberger, and A.~Wurzinger.
\newblock Post-processing subsonic flows using physics-informed neural
  networks.
\newblock In {\em 2023 AIAA AVIATION Forum}, 2022.

\bibitem{CFS}
S.~Schoder and K.~Roppert.
\newblock opencfs: Open source finite element software for coupled field
  simulation--part acoustics.
\newblock {\em arXiv preprint arXiv:2207.04443}, 2022.

\bibitem{CFSDAT}
S.~Schoder and K.~Roppert.
\newblock opencfs-data: Data pre-post-processing tool for
  opencfs--aeroacoustics source filters.
\newblock {\em arXiv preprint arXiv:2302.03637}, 2023.

\bibitem{schoder2020postprocessing2}
S.~Schoder, K.~Roppert, and M.~Kaltenbacher.
\newblock Helmholtz’s decomposition for compressible flows and its
  application to computational aeroacoustics.
\newblock {\em SN Partial Differ. Equ. Appl.}, pages 1--20, 2020.

\bibitem{schoder2020postprocessing}
S.~Schoder, K.~Roppert, and M.~Kaltenbacher.
\newblock Postprocessing of direct aeroacoustic simulations using helmholtz
  decomposition.
\newblock {\em AIAA Journal}, pages 1--9, 2020.

\bibitem{schoder2020aeroacoustic}
S.~Schoder, K.~Roppert, M.~Weitz, C.~Junger, and M.~Kaltenbacher.
\newblock Aeroacoustic source term computation based on radial basis functions.
\newblock {\em International Journal for Numerical Methods in Engineering},
  121(9):2051--2067, 2020.

\bibitem{schoder2021application}
S.~Schoder, A.~Wurzinger, C.~Junger, M.~Weitz, C.~Freidhager, K.~Roppert, and
  M.~Kaltenbacher.
\newblock Application limits of conservative source interpolation methods using
  a low mach number hybrid aeroacoustic workflow.
\newblock {\em Journal of Theoretical and Computational Acoustics},
  29(01):2050032, 2021.

\bibitem{cgal}
{The~CGAL~Project}.
\newblock {\em {CGAL} User and Reference Manual}.
\newblock {CGAL Editorial Board}, {4.10} edition, 2017.

\bibitem{tieghi2023machine}
L.~Tieghi, S.~Becker, A.~Corsini, G.~Delibra, S.~Schoder, and F.~Czwielong.
\newblock Machine-learning clustering methods applied to detection of noise
  sources in low-speed axial fan.
\newblock {\em Journal of Engineering for Gas Turbines and Power},
  145(3):031020, 2023.

\bibitem{weitz2019numerical}
M.~Weitz, S.~Schoder, and M.~Kaltenbacher.
\newblock Numerical investigation of the resonance behavior of flow-excited
  helmholtz resonators.
\newblock {\em PAMM}, 19(1):e201900033, 2019.

\end{thebibliography}

\end{document}